# Note: Observation of the angular distribution of a x-ray characteristic emission through a periodic multilayer

Philippe Jonnard,[1a] Meiyi Wu,[1] Jean-Michel André,[1] Karine Le Guen,[1] Zhanshan Wang,[2] Qiushi Huang,[2] Ian Vickridge,[3] Didier Schmaus,[3,4] Emrick Briand,[3] Sébastien Steydli,[3] and Philippe Walter,[5]

[1]*Sorbonne Université, Faculté des Sciences et Ingénierie, UMR CNRS, Laboratoire de Chimie Physique - Matière et Rayonnement (LCPMR), boîte courrier 1140, 4 place JussieuF-75252 Paris cedex 05, France*
[2]*Key Laboratory of Advanced Micro-Structured Materials MOE, Institute of Precision Optical Engineering, School of Physics Science and Engineering, Tongji University, Shanghai 200092, P. R. China*
[3]*Sorbonne Université, Faculté des Sciences et Ingénierie, UMR CNRS, Institut des NanoSciences de Paris, boîte courrier 840, 4 place Jussieu, F-75252 Paris cedex 05, France*
[4]*Université Paris Diderot-P7, F-75205 Paris cedex 13, France*
[5]*Sorbonne Université, Faculté des Sciences et Ingénierie, UMR CNRS, Laboratoire d'Archéologie Moléculaire et Structurale (LAMS), boîte courrier 225, 4 place Jussieu, F-75005 Paris, France*

We present the observation of the angular distribution of a characteristic x-ray emission through a periodic multilayer. The emission coming from the substrate on which the multilayer is deposited is used for this purpose. It is generated upon proton irradiation through the multilayer and detected with an energy sensitive CCD camera. The observed distribution in the low detection angle range presents a clear dip at a position characteristic of the emitting element. Thus, such a device can be envisaged as a spectrometer without mechanical displacement and using various ionizing sources (electrons, x-rays, ions), their incident direction being irrelevant.

About 20 years ago, André[1] proposed to use a multilayer interferential transmission plate as a soft x-ray fluorescence spectrometer. Multilayer interferential plates are periodic multilayers made of an alternation of two (or more) nanometer-thick films, where the multilayer is free standing or deposited on a membrane almost transparent to x-rays. Owing to its nanometer period, such a structure is able to diffract x-ray radiation. It was suggested to place the plate between a sample emitting some characteristic radiation to be determined and an x-ray detector. The presence of a dip in the transmittance curve of the multilayer measured as a function of the detection angle θ indicates the occurrence of a fluorescence emission. Using the Bragg law, the fluorescence wavelength λ can be determined, which makes possible the identification of the chemical element. We show the scheme of the experiment in Figure 1. We recall the Bragg law: $p\lambda = 2d \sin\theta$, where $p$ is the diffraction order, an integer, and $d$ the period of the multilayer. Using a photographic film or a CCD camera as a detector, it could be operated without scanning. Since it is based on the diffraction, by transmission through the multilayer, of the fluorescence emitted by the sample, a large variety of primary ionizing sources (electrons, x-rays, ions) can be used, the direction of the incident beam being irrelevant. No mechanical displacement of the spectrometer is needed.

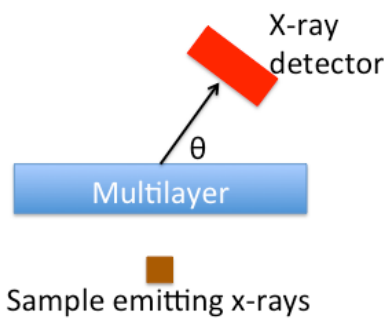

FIG. 1. Scheme of the x-ray spectrometer proposed by André on the basis of a multilayer interferential transmission plate.

Since the publication of the work of André[1], such a transmission measurement combining the angular distribution of the emitted intensity of a characteristic emission upon proton excitation and a periodic medium has never been reported to our knowledge. It is the purpose of this Note to demonstrate that this kind of work can be performed in a very efficient way with a modern

---

[a)]Author to whom correspondence should be addressed. Electronic mail: philippe.jonnard@upmc.fr.

experiment designed to study the Kossel diffraction[2–4] by a multilayer.

The present experiment is designed to measure Kossel curves, *i.e.* the angular distribution of the characteristic emissions originating from a periodic multilayer and diffracted by the multilayer itself. This last is deposited by magnetron sputtering on a silicon substrate. The diffraction element is a $[B_4C/Pd/B_4C/Y]_{x20}$ multilayer deposited on a Si substrate, the thicknesses being 1 nm for the $B_4C$ layers and 2 nm for the Pd and Y layers. The atoms of the stack are ionised by a 2 MeV proton beam at the SAFIR platform of Sorbonne Université. The protons traverse the whole multilayer and also ionise the silicon atoms of the substrate. X-rays are detected above the multilayer by an energy-dispersive CCD camera from Andor. Each CCD pixel provides an x-ray spectrum with an energy resolution of 150 eV at 5.9 keV. The CCD is set up perpendicularly to the direction of the incident proton beam. Each column of pixels of the CCD represents a detection angle so that the measurement can be made without angular scanning. The spectra of one column are summed up and we plot the intensity in a considered spectral region of interest, that of a characteristic emission, as a function of the column number, that is to say as a function of the detection angle. The angular calibration is obtained from the positions of the first and second diffraction order Kossel features of the Pd L$\alpha$ emission originating from the atoms of the multilayer, located at 2.25 and 4.16° respectively. The pressure in the experimental chamber is $10^{-3}$ Pa. The distance between the sample and the CCD is 250 mm. The proton beam current being 35 nA, about 2 h are required to perform an acquisition. More details regarding the experimental setup can be found in Ref.[4]. To draw a parallel with the spectrometer proposed by André, here the $B_4C/Pd/B_4C/Y$ system acts as the multilayer plate inserted between the detector and the substrate plays the role of the emitting sample, see Figure 1.

We show in Figure 2, the Kossel curve of the Si K$\alpha$ emission. It is very similar to the curves presented by André[1]. In this energy region the transmittance of the multilayer can reach 17%. A dip is present around 3.6° on an increasing background. The intensity variation of the background originates from the decrease of the absorption of the radiation as the detection angle increases, *i.e.* when the path travelled inside the sample is shorter. The 3.6° angle corresponds approximately to the Bragg angle calculated with the wavelength of the characteristic emission (0.713 nm[5] or 1740 eV) and the actual period of the multilayer. Owing to the broad diffraction pattern of the multilayer, the modulation of the intensity is observed over a range of about 0.5°. The shape of the observed curve can be well reproduced by a simulation, with the IMD code[6], of the transmittance of the multilayer. For this simulation no Si substrate is considered and the radiation wavelength is that of the Si K$\alpha$ emission. Let us note that the position of the dip is sensitive enough to distinguish the emission of the neighbouring elements of silicon in the period table: a dip would be observed at 4.3° for Al K$\alpha$ (1487 eV) and at 3.2° for P K$\alpha$ (2014 eV). However, owing to the large width of the diffraction pattern, the resolution is estimated to be around 0.05° in our experimental conditions, corresponding to an energy resolution of 25 eV, much smaller than the 150 eV one of the CCD. This is not yet good enough to distinguish chemical effects on photon energy, such as that occurring between the emission of silicon and silica (0.6 eV) for example which would lead to an angular shift of only 0.002°.

It would be possible to improve the spectral resolution by increasing the number of periods of the stack to decrease the width of the diffraction pattern. For example, going from 20 to 40 periods, the simulations show that it is possible to decrease the width of the Kossel feature by 0.05°. This is only a moderate decrease, about 10%, which cannot lead to a large improvement of the spectral resolution. Let us note that with 40 periods the transmittance of the stack is still a few percent, sufficient to allow measurements. An alternative to obtain a substantial improvement would be to use a multilayer etched following the profile of a lamellar grating. It has been shown that a narrowing of a factor 3 to 5 of the diffraction pattern can be obtained in this case[7,8]. Such an improvement would make possible to resolve emissions as close as a few eV. However, to obtain chemical sensitivity, such as that between silicon and silica, a further improvement would still be necessary. This could be obtained by using a crystal, provided it is sufficiently thin to be transparent to x-rays.

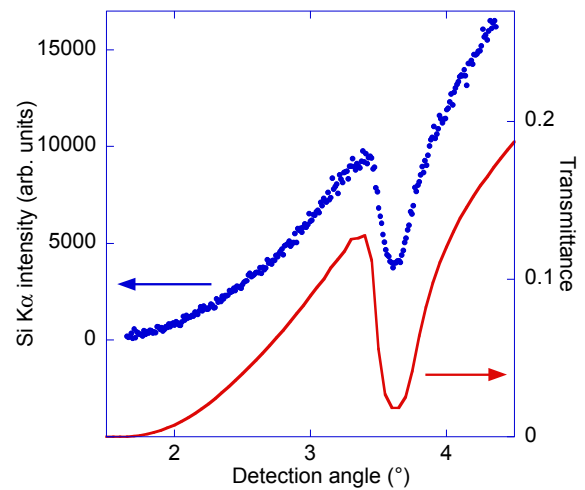

FIG. 2. Angular distribution of the transmittance of a $[B_4C/Pd/B_4C/Y]x20$ multilayer at the wavelength of the Si K$\alpha$ characteristic emission (0.713 nm) coming from the silicon substrate: solid line, simulation; dotted line, experiment. The curves are shifted vertically for sake of clarity.

Let us note that the André's experiment was performed under x-ray excitation on a synchrotron beamline[1]. In this condition, a low intensity background extending toward the low angles was present. It has now been demonstrated that such an experiment can be performed using proton excitation[4]. The interest of using an ion beam to generate the ionizations in the material under study is to work with a very low intensity of the primary beam Bremsstrahlung[9], leading to high signal to noise ratios. This would allow analysing the characteristic emissions of elements present in low concentration, such as impurities or dopants. A further advantage of using charged particles is that they may be focused and scanned to obtain micronmeter resolution maps of inhomogeneous samples. The nature and energy of the ions may also be selected to optimise the ionization cross section and analysed depth according to the sample thickness and composition. However, probing insulating materials with an ion beam can be difficult because of possible charge effects, whereas this type of problem does not exist upon photon excitation (x-ray fluorescence, XRF), which is a photon-in photon-out method. Let us also note that XRF experiments can be performed on synchrotron facilities and benefit from the very large incident photon flux and thus compensate for the small solid angle of detection needed to obtain the desired angular resolution.

Finally, the use of an energy-dispersive CCD camera[10–13] as a x-ray detector enables simultaneous observation of the angular and spectral distributions of the emitted radiation, thus working at the same time in both wavelength dispersive (WDS) and energy dispersive (EDS) spectroscopic modes. This reduces the acquisition time necessary to obtain a workable curve to a few minutes, makes it possible to observe and detect simultaneously the presence of more than one element in the emitting sample and allows improving the spectral resolution provided by the CCD camera.